# Condensation of holes into antiferromagnetic droplets in the organic semiconductor (DOEO)$_4$[HgBr$_4$]·TCE

O.V. Koplak[1*], A. Chernenkaya[3, 4], K. Medjanik[3], A. Brambilla[1], A. Gloskovskii[5], A. Calloni[1], G. Schönhense[3, a)], F. Ciccacci[1], R.B. Morgunov[2, 6]

[1] *Dipartimento di Fisica, Politecnico di Milano, 20133, Milano, Italy*

[2] *Institute of Problems of Chemical Physics, 142432 Chernogolovka, Moscow, Russia*

[3] *Institut für Physik, Johannes Gutenberg-Universität, D-55099 Mainz, Germany*

[4] *Graduate School Materials Science in Mainz, Staudingerweg 9, 55128, Mainz, Germany*

[5] *Deutsches Elektronen-Synchrotron DESY, Notkestr. 85, D-22607 Hamburg, Germany*

[6] *Sholokhov Moscow State University for the Humanities, Moscow, Russia*

Changes of the electronic structure accompanied by charge localization and a transition to an antiferromagnetic ground state were observed in the (DOEO)$_4$[HgBr$_4$]·TCE organic semiconductor. Localization starts in the region of about 150 K and the antiferromagnetic state occurs below 60 K. The magnetic moment of the crystal contains contributions of antiferromagnetic inclusions (droplets), individual paramagnetic centers formed by localized holes and free charge carriers at 2 K. Two types of inclusions of 100-400 nm and 2-5 nm sizes were revealed by transmission electron microscopy. Studying the symmetry of the antiferromagnetic droplets (100-400 nm inclusions) and individual localized holes by electron spin resonance (ESR) revealed fingerprints of the antiferromagnetic resonance spectra of the spin correlated droplets as well as paramagnetic resonance spectra of the individual localized charge carriers. Photoelectron spectroscopy in the VUV, soft and hard X-ray range shows temperature-dependent effects upon crossing the critical temperature. The substantially different probing depths of soft and hard X-ray photoelectron spectroscopy yield information on the surface termination. The combined investigation using soft and hard X-ray photons to study the same sample results in details of electronic structure including structural aspects at the surface.

---

[a)] Author to whom correspondence should be addressed. Electronic mail: schoenhe@uni-mainz.de.

[*] On leave from.

**I. INTRODUCTION:**

Organic conductors are promising candidates for future molecular-scale memory applications. Reviews about the important practical applications and novelty of such materials were presented in [1-3]. The possibilities of molecular design through chemical synthesis and of miniaturizing dimensions are attractive properties of the molecular based conductors. Individual or small packets of molecules mounted within addressable cells conduct and switch electrical currents, and provide electrical bits of information [3]. Advantages of organic and polymer memories also include simplicity in device structure, good scalability, low fabrication costs, low-power operation, multiple properties interplay, and large capacity for data storage. Organic metals and semiconductors are strongly-correlated Fermi liquid systems manifesting unique physical properties such as charge and spin waves, unusual electrical and optical properties [4, 5]. Strong electron-electron correlations and electron – phonon interactions appearing due to low dimensional structure of the layered organic conductors cause unusual ground states of the crystals possessing a very rich spectrum of instabilities [4, 5]. Instabilities lead to metal–insulator or metal–semiconductor transitions controlled by the amount of Fermi surface involved in the electron–hole condensation.

Most of the organic conductors are layered two dimensional (2D) crystals whose symmetry contains an inversion center. The inversion symmetry forbids spin-orbit contribution like Aharonov- Bohm or Rashba-Bychkov effects. This paper is devoted to the quasi-2D organic conductor containing the asymmetrical DOEO molecule (1,4-(dioxandiil-2,3-dithio) ethylenedithiotetrathiafulvalene) packed in two dimensional layers. We have studied $(DOEO)_4[HgBr_4] \cdot TCE$ (where TCE is 1,1,2-trichloroethane) single crystals (Fig. 1). The DOEO based crystals are layered systems attractive because conductive sheets of monomolecular thickness are separated by relatively large nonconductive zones of the anionic layer. Analogously to the family of BEDT-TTF cationic radical salts, the charge transfer from the DOEO layer amounts to 1 electron per molecular dimer. Synthesis, crystal structure, electrical and optical properties of the $(DOEO)_4[HgBr_4] \cdot TCE$ semiconductors were described in detail [6-9]. Raman scattering investigations showed that the crystals are quasi-one dimensional (1D) conductors down to 140 K, whereas below this temperature the conductivity behavior becomes similar to the one typical for quasi-2D compounds [6]. Upon cooling the resistivity of the crystals increases until a maximum at ~ 140 K corresponding to a transition to the (anomalous) metallic state existing in the 70-140 K range. Below about 60 K a sharp increase of the resistivity was observed [7]. The highest occupied molecular orbital (HOMO) of the DOEO molecule of π character constitutes a 2D conduction band and there is energy gap forming by two coupled DOEO molecules in the HOMO band. Therefore, the HOMO band is expected to be nominally quarter-filled.

The aims of this paper are: 1) identification of the particles contributing to magnetic and electrical properties in $(DOEO)_4[HgBr_4]\cdot TCE$ single crystals; 2) studying the local symmetry of the crystal field in the region of the charge carrier localization; 3) identification of energy structure and chemical composition of the crystals by X-ray photoelectron spectroscopy (XPS) and hard X-ray photoelectron spectroscopy (HAXPES) in complicated multiphase system.

## II. EXPERIMENTS

Single crystals $(DOEO)_4[HgBr_4]\cdot TCE$ were of plate shape of ~1 mm$^2$ sizes. The structure of these crystals consists of alternating organic and inorganic layers (Fig. 1 b). The non-conducting inorganic layers are formed by $(HgBr_4)_2$ anions and TCE solvent molecules. The organic layers are built by dimerized DOEO molecular stacks (Fig. 1 a).

To investigate the high frequency spin dynamics and to separate contributions of particles of various types to the magnetic susceptibility of the crystal a Jeol JES FA200 electron spin resonance (ESR) spectrometer was used. The spectrometer operates in the X–band frequency range (9.013 GHz). The modulation frequency was 100 kHz, scanning range of the magnetic field was in the 0–16 kOe range. The temperature was varied in the 4 - 300 K range with a relative accuracy of ± 5% in an ESR 900 Oxford Instruments cryostat. The ESR spectra were recorded as dependences of the first derivative of microwave power absorption on the magnetic field, d$P$/d$H$. For measurements of the angular dependence the crystal was mounted on the sample holder with an absolute accuracy of ± 5$^0$. Rotation of the sample in the cavity was carried out by a homemade goniometer with a relative error of ± 1$^0$.

Ultraviolet and X-ray photoelectron spectroscopies (UPS and XPS, respectively) were used to determine the electronic and chemical properties of the compound. The XPS analysis utilized Mg $K\alpha$ radiation ($h\nu$=1253.6 eV), while hard X-rays were produced by the storage ring PETRA, Hamburg, thus providing two different scenarios of the probing depth. The combination of core- and valence photoemission spectroscopy is generally necessary for providing a full picture of the electronic structure. The XPS/UPS and HAXPES measurements were performed with two hemispherical analyzers (type SPECS Phoibos) for the standard and high-energy range, respectively. UPS employed a He discharge lamp (He I at 21.2 eV) for excitation. The full width at half maximum (FWHM) resolution was about 0.9 eV, 0.4 eV and 0.3 eV for XPS, HAXPES and UPS, respectively. All spectra were acquired at normal emission. The UHV sample manipulators were equipped with liquid helium cryostats, thus providing temperatures in the ca. 15-300 K range.

The HAXPES measurements were carried out at the P09 HAXPES end station of PETRA III (Hamburg) at photon energy of 6000.5 eV. Due to the large inelastic mean free path of the fast electrons the probing depth was about 8 - 10 nm and thus surface contaminations or the termination of the surface play only a minor role. Comparison of XPS and HAXPES spectra thus allows drawing conclusions about the surface termination of the as-grown crystals. Due to this large probing depth, more than one layer of DOEO molecules contributes to the photoelectron signal in the HAXPES measurements. So the interlayer conductivity might become visible in this spectral range. For maximum sensitivity and minimum radiation damage the energy analyzer (type SPECS Phoibos 225 HV) was combined with a delay-line detector (type Surface Concept). For the present measurements the overall energy resolution (electrons and photons) was 440 meV. The beam was focused down to 0.2 mm × 0.4 mm spot size. The spectra were recorded in normal emission geometry at nearly grazing photon incidence ($\approx$ 2°).

Finally, the microstructure of the sample has been studied by transmission electron microscopy (TEM) using a Jeol JEM-2100 high resolution microscope.

## III. RESULTS AND DISCUSSION

### A. Magnetic properties

The ESR spectroscopy can indicate the localization of charge carriers via characteristic changes in the ESR spectrum, in particular spectral shape and symmetry. ESR gives information on the surroundings of the localized paramagnetic centers. For that reason ESR spectra of a $(DOEO)_4[HgBr_4] \cdot TCE$ single crystal were recorded (Fig. 2). The ESR spectrum of $(DOEO)_4[HgBr_4] \cdot TCE$ crystal contains a single line of Lorentz shape in the 4 – 300 K temperature range (Fig. 2 a). At higher temperatures, 70 – 300 K, this line possesses an axial angular dependence inherent to delocalized charge carriers in the organic conductors (Fig. 2 c).

Below $T = 70$ K the ESR spectrum is split into two anisotropic lines *1* and *2* (Fig. 2 b). The narrow line *1* (linewidth ~ 0.5 - 4 mT) which was also observed at high temperatures lies on the low-field wing of the wide line *2* (linewidth ~ 2 - 8 mT). Decomposition of the spectra recorded in different orientations into two Lorentz lines results in angular dependences of the *g*-factors of lines *1* and *2* (Fig. 2 c). Obvious differences of the angular dependences at 300 K and 5 K as well as pronounced differences between angular dependences of lines *1* and *2* indicate the presence of manifold paramagnetic centers corresponding to different crystal fields. In Section D evidence will be presented that low temperature centers contribute to the ESR spectra by providing localized charge carriers.

A sharp increase of the sample resistivity below 70 K was observed in [6-9] and confirmed in special measurements for the crystal subjected to ESR studies. The room temperature dc conductivity of the single crystal measured along the *a*-direction within the *ab* plane was $R_{300}$ = 5 S/cm, while the perpendicular one was smaller by almost one order of magnitude. Upon cooling, the resistivity increases until a maximum is reached around 140 K. Below this maximum the resistivity curve becomes metallic-type and finally undergoes an insulating transition at about 50 K. Detailed investigation of the electrical conductivity was not general aim of the present paper and it was explored just to confirm the similarity of our studied crystals with the ones studied in [6-9]. The resistivity of the sample *R* depends on the amount of delocalized charge carriers $N_{del}$: $1/R = N_{del}\mu e k$ ($\mu$ is mobility of the charge carriers, *k* is a geometric factor, *e* is the hole charge). Delocalized charge carriers (holes) do not strongly contribute to the magnetic moment of the sample *M*, because just a small part of them with energies close to the Fermi level $E_F$ (softening zone ~ $4k_BT$) contributes to *M*. The number of localized holes $N_{loc} = (N - N_{del}) = N - 1/R$ obeying Curie law should mainly contribute to *M*. The temperature dependence of $N_{loc}$ (presented in arbitrary units) is shown in Fig. 3 a as well as the temperature dependence of magnetization *M*. Good correlation between $N_{loc}(T)$ and $M(T)$ dependences indicates the dominant contribution of the localized holes to the magnetic moment of the sample. The slight discrepancy between these curves may be explained by experimental details. A single crystal was used for the measurement of the resistivity while the measurement of the temperature dependence of the magnetic moment by a SQUID magnetometer requires a large amount of sample material. To obtain the *M(T)* dependence about 8 mg of powder containing small crystals of ~ 0.1 mm sizes was used. Since the maximum of the $N_{loc}(T)$ curve varied from sample to sample, the magnetization corresponds to the average amplitude which is smaller than the one observed in individual samples. Provided a sufficient amount of material, magnetometry is a good tool for estimation of the concentration of localized holes.

The main question implied to answer by magnetometry concerns the nature of the low temperature magnetic state of the localized carriers in the crystals. Measurements of the magnetic field dependence of the magnetization were performed to clarify this problem (Fig. 3 b). One can expect a Brillouin field dependence of the magnetization in case if a paramagnetic state of the crystal is present. In spite of this expectation we have found a superposition of a Brillouin function (~ 70%) with a linear field dependence characteristic for an antiferromagnetic contribution (~ 30%). The crystal structure and low concentration of holes (one hole per molecular dimer [6-9]) give no chance to observe long range intermolecular exchange interaction because intermolecular distances exceed ~ 10 Å being longer than any reasonable exchange radius ~ 1 Å. Thus, the observed linear part of the magnetic field dependence *M(H)* implies the existence of magnetic centers with locally short distances between them ~1Å to provide direct exchange interaction. There are no such short distances in the initial crystal structure. Nevertheless, localization of the charge carriers leads to appearance of individual paramagnetic centers at intermediate temperatures (70-140 K) and antiferromagnetic groups ("droplets") of holes at low temperatures below about 60 K.

Condensation of holes into antiferromagnetic "droplets" requires additional confirmations which will be given in Section D. Nevertheless the magnetic data present a background for this statement. An antiferromagnetic contribution of the "droplets" gives a reason to discuss ESR line *2* as antiferromagnetic resonance. As an independent confirmation of this view point continuous decreasing temperature dependence of the integral intensity of line *2* can be considered [10].

**B. X-ray photoelectron spectroscopy**

XPS is very sensitive to the chemical composition and environment of the elements in a material in a probing depth typically lower than a few nm. In XPS spectra, each element will give rise to a characteristic set of peaks in the photoelectron spectrum at kinetic energies determined by the photon energy and the respective binding energies. The presence of peaks at particular energies therefore indicates the presence of a specific element in the sample under study. Furthermore, the intensity of the peaks is related to the concentration of the element within the sampled region. Thus, the technique provides a quantitative analysis of the surface composition. Arising out of the fact that the transfer of charge from the valence orbitals, typically involved in chemical bonding, is reflected in the measured binding energies of the core levels of the atoms, it is possible to distinguish the signals originating from atoms in different chemical surrounding.

The hard X-ray variant of XPS (HAXPES) provides the same information but for a substantially larger probing depth of 10-15 nm, corresponding to 7-10 DOEO layers of the title compound. HAXPES thus makes it possible to investigate the electronic structure in the bulk of the material. Furthermore, the comparison of bulk and surface photoelectron spectra gives access to the surface termination. A combined study using soft and hard X-ray photons on the same sample allows to discuss the electronic structure taking into account different structural aspects at the surface.

XPS and HAXPES spectra of $(DOEO)_4[HgBr_4]\cdot TCE$ crystals were taken in the temperature range 45 K – 300 K. The survey XPS spectrum of the DOEO salt shows all the elements present in the samples (Fig. 4 a). The spectrum contains the carbon, oxygen and sulfur signals corresponding to the conductive DOEO stack layers as well as mercury and bromine present in the anion sublattice and chlorine from the TCE solvent.

A HAXPES survey spectrum of the shallow core levels, recorded at room temperature, is shown in Fig. 5 a. Like in the XPS case, the spectrum contains the signals of all DOEO components, however, with substantially different intensity ratios. We observe chlorine of the solvent molecules embedded in the anionic layer. At room temperature the signals are well resolved and non-split.

Let us now compare the relative intensities of the anionic components mercury and bromine. Their intensities in the HAXPES survey spectrum are much higher than those for the surface-sensitive XPS in comparison to signals of C 1s and O 1s that have the largest intensities in the spectra (see Fig. 4 a and 5 a). The XPS detects higher S/Hg and S/Br ratios than expected from the composition, suggesting that the surface of the as-grown crystals is mainly terminated by the layer of DOEO molecules. HAXPES, with its large probing depth, gives spectroscopic access to the buried anionic layers. This general finding of termination of the surface by the DOEO layer is in agreement with results from scanning tunneling spectroscopy (STS) on the related compound $(BEDT-TTF)_2Cu[N(CN)_2]Br$, where tunneling spectra of as-grown crystals gave direct evidence of the states in the cationic layers in the superconducting phase [11].

Along with the discussion of photoemission spectra we should mention two main phenomena that could contribute to the experimental HAXPES and XPS spectra besides the true signal of the organic material: radiation damage and charging of the crystals or parts of them. In order to avoid the effect of radiation damage, several crystals from different batches were measured in parallel at each temperature in our experiments. Furthermore, the X-ray beam was shifted to a new position at the sample before every new scan. Charging often occurs during measurements on insulating materials. The inhomogeneous charge distribution in the beam spot normally broadens the XPS lines and shifts the whole spectrum to lower kinetic energies. In XPS, we observe the charging effect very close to the surface. However, in HAXPES the large mean free path of several DOEO layers can lead to inhomogeneous, layer-wise charging as discussed in [12]. The present results also indicate a charging contribution that could not be avoided. The measurements were reproduced on several samples from different batches and spectra were taken both during cooling and during warming up from low temperatures. The trend for all samples was similar. For further evaluation the S 2s, Hg 4d and Cl 2p HAXPES signals were used as a probe for the DOEO molecule, the anion and the solvent component, respectively (Fig. 5 b). These lines are more suitable for the quantitate discussion, because their intensities are larger and the component structure is more pronounced than for others due to photoemission cross sections that decrease with increasing of kinetic energy.

The Hg 4f XPS spectrum (Fig. 4 b) displays three peaks, with the Hg 4 $f_{7/2}$ and 4 $f_{5/2}$ spin-orbit doublet appearing at 100.5 and 104.5 eV, respectively. The third peak showing up in the Hg 4f region, located at 102.2 eV, can be associated to Si oxides, whose presence is likely due to surface contaminants. In the HAXPES spectra of Hg we do not observe a splitting at room temperature, because the surface contribution is very small in comparison to the bulk contribution of several anionic layers. Furthermore, we do not observe a signal from Si in the bulk.

The S 2p XPS level (see Fig. 4 c) is the convolution of a spin-orbit split doublet in which the binding energies of the S $2p_{3/2}$ and the S $2p_{1/2}$ core levels are 164.2 eV and 165.3 eV, respectively, and of another doublet, characterized by a significantly lower intensity, with binding energies of 165.4 eV and 166.5 eV for the $2p_{3/2}$ and $2p_{1/2}$ components, respectively. The presence of two S doublets can be associated either with photoemission from nonequivalent S sites in the DOEO molecule, or to the presence of DOEO molecules with a different charge states. Similar observations are reported in the literature for the photoemission from organic salts (see e.g. [13] and references therein).

The Hg 4d and Cl 2p HAXPES spectra of the anionic layer components have the same tendency of development upon variation of the temperature. The spectra at 200 K consist of single lines. Normally, the Cl 2p has two contributions located at about 200 and 202 eV of binding energy. In the present experiment we could not resolve this difference probably due to a highly disordered anion-solvent system that was described in the structural investigation [6]. Cooling down to 130 K does not change the Hg 4d and Cl 2p spectra, although there is a shift of 1.5 eV to lower kinetic energies. Upon further cooling down to 70 K additional lines in Cl 2p, Hg $4d_{3/2}$ and Hg $4d_{5/2}$ appear and the spectra are shifted further by 2.2 eV in the same direction. The S 2s signal behaves a bit differently, because it exhibits a low-energy satellite signal, shifted by about 4eV already at 200 K. At 130 K this satellite became significantly stronger and at 70 K it shows the same large shift as the Cl 2p signal.

The hypothesis about a transition from 1D electrical conductivity (at high temperatures) to 2D conductivity (below 140 K) proposed in [9] stimulated us to study the temperature dependence of the electrical anisotropy. In order to quantify the electrical anisotropy we introduce the relative difference of the "in plane" $R_{\parallel}$ and "out-of-plane" $R_{\perp}$ resistivities:

$$\Delta R/R = (R_{\perp} - R_{\parallel})/R_{\parallel} \qquad (1)$$

Two facts become obvious from the comparison of temperature dependencies of $\Delta R/R$ and the XPS shift, both plotted in Fig. 6. Cooling of the crystal continuously increases the anisotropy of the electrical conductivity towards two- dimensional behaviour. This continuous "transition" of the electrical dimensionality accompanied with localization of the individual holes leads to dielectric polarization of the sample. The additional evidence of localization (except early described resistance behaviour [6-9]) is a significant shift towards higher binding energies of the core shell XPS lines progressively down to 140 K. The temperature dependence of the electrical anisotropy parameter (1) is in good correlation with the temperature dependence of the level shifts in the XPS spectra (Fig. 6). In the DOEO layer the conductivity originates from transfer of holes between dimers through short contacts between "tail-to head" molecules whose ends contain sulphur atoms. Further, also the intensity ratio of the HAXPES signals (i.e. the ratio of the area of the shifted line and the total area of shifted and non-shifted line) also follows this trend. The net charge accumulated in the near-surface region in the steady state equilibrium during irradiation with the hard X-rays should also depend on the interlayer conductivity.

### C. Ultraviolet photoelectron spectroscopy

UPS probes the single-particle spectral function and is therefore a useful tool to investigate highly correlated systems. According to the atomic subshell photoionization cross sections calculated in [14], the contributions of the C 2p and S 3p orbitals are very large for the He I photon energy whereas those of Hg 5d is rather small because of the small cross sections and small atomic ratios [15]. The general structure of He I-excited UPS spectra is characterized by a strongly increasing background intensity of inelastically scattered photoelectrons (i.e., the so-called secondary electrons). Quantitative information on the electronic level alignment at the DOEO salt is provided by means of UPS, which may also allow for an estimate of the surface work function. The surface work function $\Phi$ of DOEO was estimated from the onset of the secondary electrons, to be 4.1±0.2 eV. The work function of a semiconductor, however, is not solely determined by the ionization energy but also by the position of the Fermi level at the surface, which may be shifted by the occupation of surface states in the forbidden gap. From the estimated value of $\Phi$ and assuming an ionization energy $E_I$ = 4.8 eV for the neutral DOEO molecules [7], it is possible to give an estimate of the expected position for the HOMO level, at a binding energy of 0.7±0.2 eV.

Figure 7 shows a UPS spectrum of the $(DOEO)_4[HgBr_4]\cdot TCE$ crystal taken at 300 K in the whole valence-band region. Such a spectrum is characterized by a slow increase in the 2 eV to 4 eV region, where S 3p states are expected to give the dominant contribution [15, 16]. Then there is a strong rise towards a plateau that encompasses the 6 eV to 8 eV binding energy region. Here, one should expect a contribution also from Hg 5d, C 2p and S 3s states [15, 16].

Finally, the inset of Fig. 7 is an enlargement of the He I spectrum around the Fermi level. Here, the onset of the electron counts suggests that our estimate for the HOMO positioning is reasonable.

As for XPS, also the onset of electron counts in UPS spectra is seen to shift towards higher binding energies when decreasing the sample temperature. The temperature dependence of such a shift is shown in Fig. 8. In the UPS case, the shifting towards higher BE is also accompanied by distortions in the spectra, which ultimately hinders a clear comparison between the UPS line shapes recorded at the different temperatures. For this reason, the behavior described by Fig. 8, which noticeably shows a slight discontinuity around 150 K, can only give a qualitative evaluation of the temperature dependence of the UPS spectra. Such dependence seems anyway to be in fair agreement with that reported above for the XPS case.

Photoemission and electronic structure near the Fermi level ($E_F$) in similar TTF-based crystals was studied by R. Liu et al. [16]. Significant photoelectron-emission intensities are observed in the region close to $E_F$, in those salts showing a metallic behavior. However, the spectrum of Fig. 7 displays a broad onset from the Fermi level with emission intensity below 0.5 eV from $E_F$, where the metallic bands are expected to be located, being near zero. This result is in contrast to a sharp Fermi edge observed in most of the three-dimensional (3D) and two-dimensional (2D) metallic systems. Electron-electron and electron-phonon interactions are considered as the likely causes for the absence of a sharp Fermi edge. However, if we think at the DOEO salt as to a semiconductor, the negligible intensity seen at $E_F$ should be due to the presence of a gap. If the salt would behave as a metal, then the effect of correlations may be invoked to explain the absence of a sharp Fermi edge.

The vibrational structure of the $DOEO^0$ neutral molecule, the $DOEO^+$ cation, and the TCE solvent molecule were described using the DFT method (density functional theory) [7]. For fully optimized $DOEO^0$ the skeleton is nonplanar, whereas for $DOEO^+$ it is almost planar (except for the outer ethylene groups). Computation of the electronic structure of $DOEO^0$ and $DOEO^+$ was performed using quantum-chemical calculations at different levels of theory and gives HOMO energy values of 4.8 eV and 8.5 eV as referred to the vacuum level (corresponding therefore to 0.7 eV and 4.4 eV in our reference system, respectively) for the neutral and charged $DOEO^+$ molecule, respectively. The theoretical LUMO energy (referred to the vacuum level) is about 1.3 eV for $DOEO^0$ and 7.2 eV for $DOEO^+$ [7]. If the conduction band originates from the HOMO in the $(DOEO)_4[HgBr_4]\cdot TCE$ salt we would expect to have delocalized conduction electrons parallel to the stacking axis. The direct overlapping of the planar part of the DOEO molecule with an interplanar distance of 3.57 Å enables intermolecular electron transfer through the π orbitals.

**D. Transmission electron microscopy and "droplets" contribution to the electrical and magnetic properties**

Different resolution HRTEM images of a $(DOEO)_4[HgBr_4] \cdot TCE$ single crystal are shown in Fig. 9 a, b. One can observe two types of inhomogeneous features that are: 1) "droplets" of 100-400 nm diameters occupied about 20-30 % of the crystal bulk (Fig. 9 a), 2) a mosaic structure containing 5-7 nm disoriented blocks of atomic planes separated from each other by 2.66 Å (Fig. 9 b). These heterogeneities possess crystal structure being of the same symmetry as the main crystal structure. Heterogeneities were revealed at room temperature but of course it is likely that they persist at low temperatures. It is well known that instability of the crystal lattice as well as energy balance can result in phase separation. It is reasonable to suppose that the "droplet" inclusions as well as the mosaic substructure provoke charge carriers localization providing channel for holes runoff. Coincidence of the fraction of antiferromagnetic material estimated from magnetic moment field dependence with the fraction occupied by the "droplets" allows us to suggest that charge carriers are partially localized in the "droplets" at 2 K. To resume obtained results the main spin carriers contributing to the ESR and SQUID signals were sketched in Fig. 9 c. They are free holes moving along trajectories, localized individual holes forming paramagnetic centers and antiferromagnetic droplets. One should expect Schottky barriers forming on the droplet interface and providing Coulomb repulsion of the charge carriers balanced with equilibrium concentration of the holes in the crystal bulk. This is a way to explain non-monotonous temperature dependencies of magnetization and electrical characteristics (Fig. 3 a).

As is well known, two different materials (bulk and droplets) generally have different work functions. When they are attached to each other, the energy levels are modified around the interface by transferring electrons to match their Fermi levels. This is often called band bending and forms a Schottky barrier at the metal-insulator interface. The electronic state in the close vicinity of the interface can be modified physically and/or chemically and depends sensitively on the two materials. Nevertheless, the overall band structure is governed by the long-range Coulomb interaction, i.e., by the Poisson equation if the band structure can be continuously treated. Competition between temperature dependencies of the Fermi level position and height of the Schottky barrier results in a kink on the $M(T)$ and $R(T)$ dependencies as it was described in [17] for semiconductor heterojunctions.

**IV. CONCLUSIONS**

At low temperature an antiferromagnetic phase of localized holes coexists with paramagnetic individual localized holes in (DOEO)$_4$[HgBr$_4$]·TCE. The symmetry of the "islands" of the antiferromagnetic phase was found to be different in comparison with the symmetry of the ESR spectra of free charge carriers.

In the temperature region of 120-160 K the XPS spectra reveal characteristic level shifts of the order of 400 meV and the HAXPES spectra exhibit strongly shifted signals due to the buildup of non-equilibrium charge distribution in the near surface region. The temperature dependences of the XPS level shift and the intensity ratio of the shifted signal versus the unshifted signals in HAXPES spectra agree well with the conductivity anisotropy parameter $(R_\perp - R_\parallel)/R_\parallel$.

The comparison of XPS and HAXPES spectra gives information on the surface termination. XPS (probing depth about 1nm) sees the surface and topmost DOEO layer, whereas HAXPES with its large probing depth of 10-15 nm gives spectroscopic access to many buried cationic and anionic layers. The comparison indicates that the surface of the as-grown crystals is mainly terminated by the layer of DOEO molecules.

Continuous localization of holes results in non-monotonous variations of the resistivity and magnetization of the samples corresponding to a competition between free charge carriers, individual localized holes and condensed droplets containing large amounts of spins. The obtained data can be used to explain extra ordinary electrical and magnetic properties of the (DOEO)$_4$[HgBr$_4$]·TCE crystals. Appearance of an antiferromagnetic phase below 60 K in a non-magnetic compound not containing metal atoms is an evidence of the strong electron-electron correlation in this two dimensional system.

## ACKNOWLEDGMENTS

O.V.K. thanks Cariplo Foundation, R.B.M. thanks the Russian Foundation for Basic Research (Project 13-07-12027) for financial support. Financial support by DFG (Transregio SFB TR49), BMBF (05K13UM4) and Graduate School of Excellence MAINZ is gratefully acknowledged. The authors are grateful to I. Khodos for assistance in HRTEM measurements, to H. J. Elmers for fruitful discussions about the magnetic contributions and to W. Drube with team for support during the measurements at beamline P09 (PETRA III at DESY, Hamburg). A.C. research is founded by Fondazione Cariplo (2012-09-04 SECARS project).

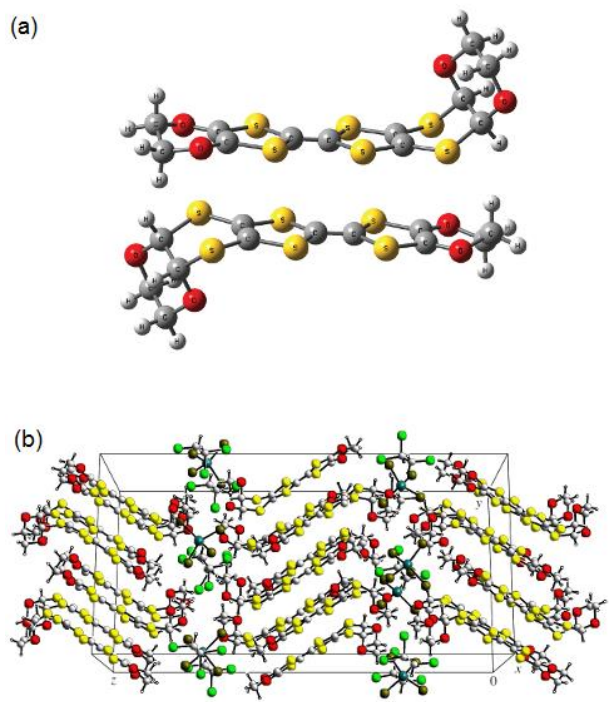

FIG. 1. Molecular dimer DOEO (a) and (DOEO)$_4$[HgBr$_4$]·TCE crystal (b) structures revealed in Refs [6-9].

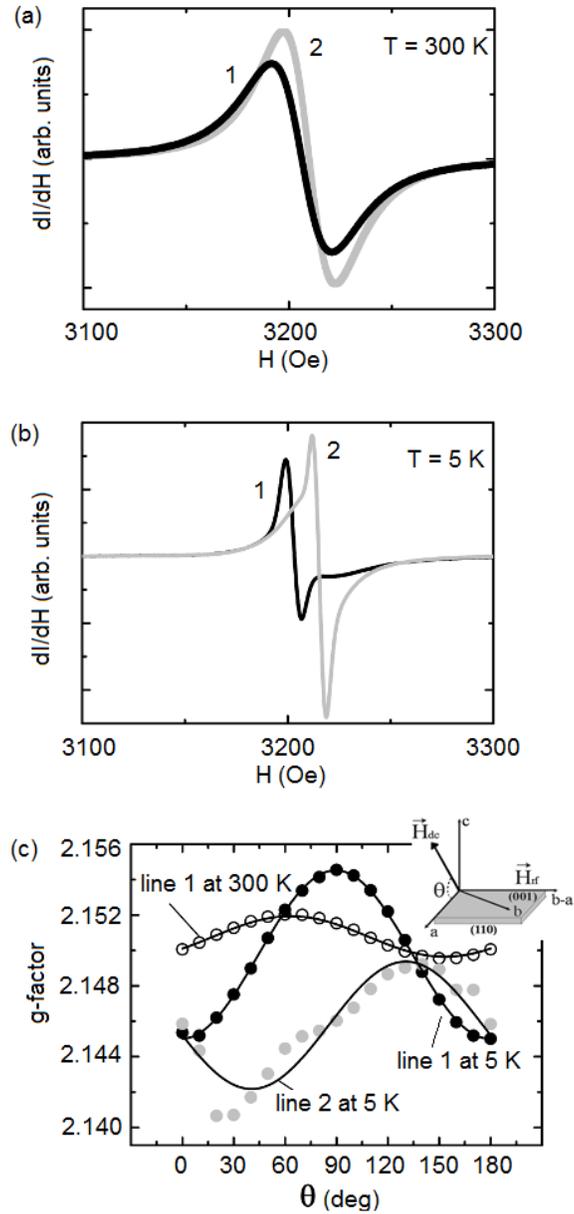

FIG. 2. (a, b) ESR spectra of a (DOEO)$_4$[HgBr$_4$]·TCE single crystal at $T = 300$ K (a) and 5 K (b). In both panels the angle between the magnetic field and the *ab* crystal plane is $\theta = 70°$ (gray lines) and $\theta = 150°$ (black lines). (c) The "out-of-plane" angular dependence of *g*-factors of a (DOEO)$_4$[HgBr$_4$]·TCE single crystal obtained from the ESR spectra at $T = 5$ K for line *1* (black symbols) and line *2* (gray symbols). Open symbols give the angular dependence of line *1* at $T = 300$ K. The inset shows the orientation of the microwave and static magnetic field of the spectrometer with respect to the crystal, as well as the direction of sample rotation. The solid lines are approximations of the angular dependences by the spin Hamiltonian corresponding to axial symmetry. [10]

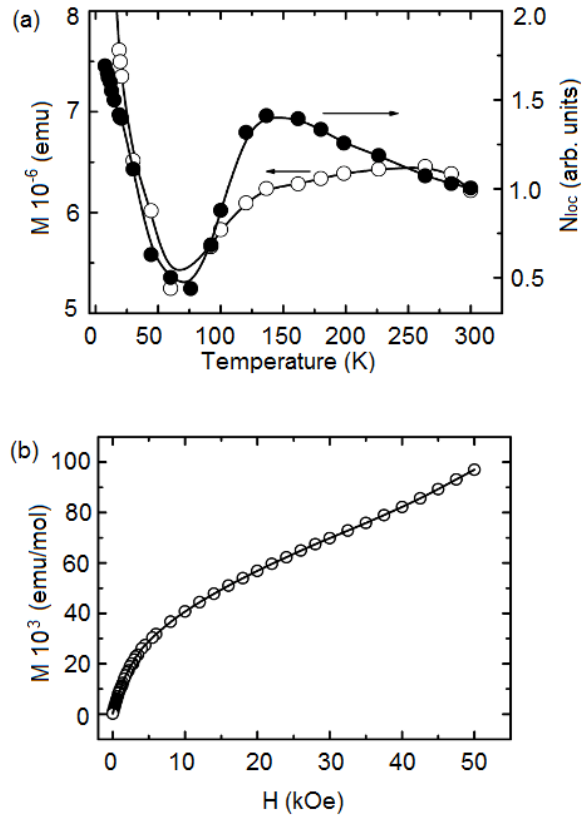

FIG.3. (a) Temperature dependencies of the magnetic moment of the sample $M$ in constant magnetic field 1 kOe and number of localized charge carriers $N_{loc}$ recalculated from resistivity, (b) Field dependence of magnetic moment of $(DOEO)_4[HgBr_4]\cdot TCE$ single crystal at 2 K. The solid line shows the approximation of the magnetic moment with sum of the Brillouin function with $S = \frac{1}{2}$ and linear dependence corresponding to antiferromagnetic contribution.

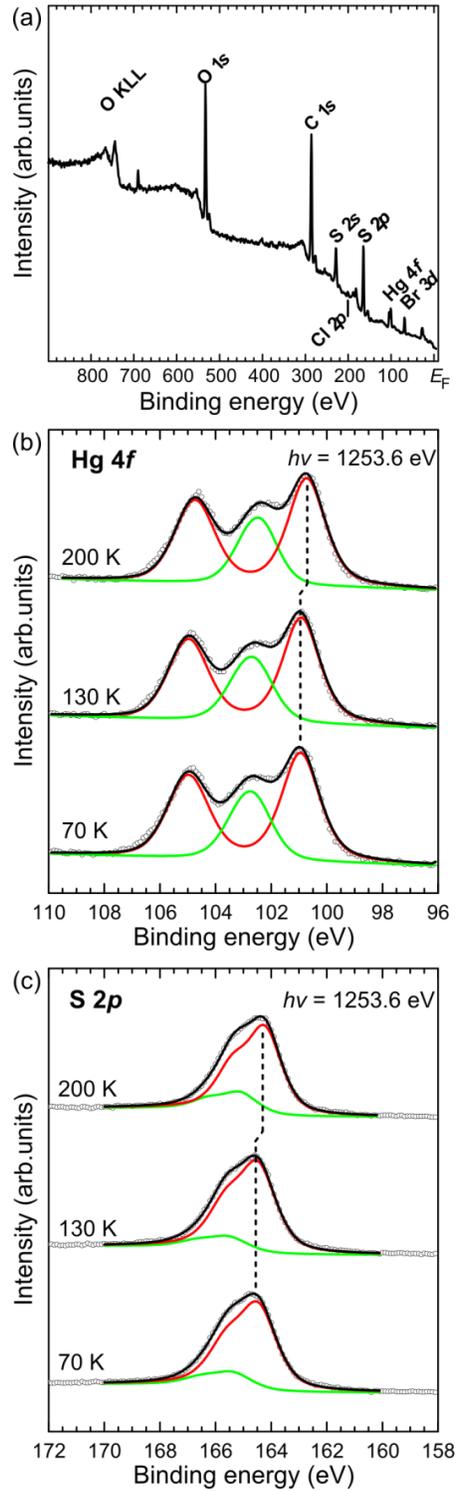

FIG. 4. (a) Survey XPS spectrum of the surface of the (DOEO)$_4$[HgBr$_4$]·TCE crystal, (b) an example of the temperature behavior of Hg 4f core-level XPS spectra at 45 K (open circles), 100 K (gray circles) and 250 K (black circles, solid lines show deconvolution), (c) example of the S 2p core level spectrum decomposition at 45 K (open circles) and 250 K (black circles), solid lines show deconvolution. In both cases the sum spectrum of the deconvoluted peaks agrees perfectly with experiment.

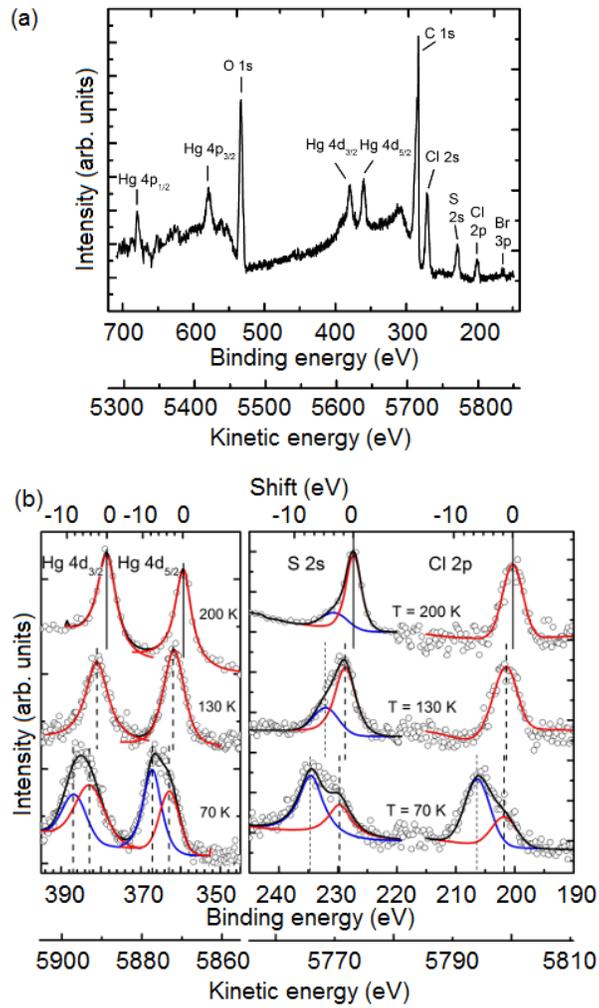

FIG. 5. (a) Survey HAXPES core level spectrum of $(DOEO)_4[HgBr_4]\cdot TCE$ taken at room temperature; (b) Mercury 4d (left panel), sulfur 2s and chlorine 2p (right panel) HAXPES core-level spectra for three temperatures of $(DOEO)_4[HgBr_4]\cdot TCE$: $T = 200$ (top), 130 (center) and 70 K (bottom). Circles show the experimental results, solid lines represent the deconvolution of the spectra by a least squares fit. Vertical lines show the center positions of the deconvoluted peaks.

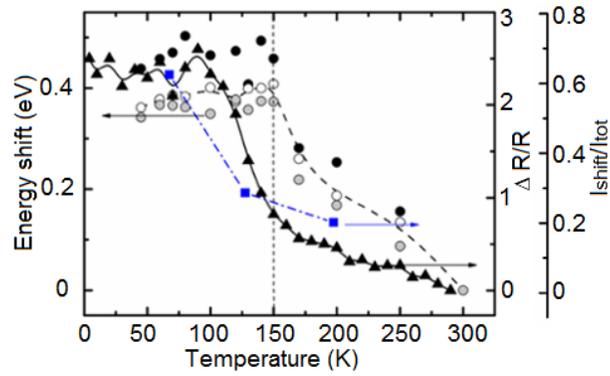

FIG. 6. The temperature dependencies of the shift of core levels of Hg (black circles), S (gray circles) and Br (white circles) in (DOEO)$_4$[HgBr$_4$]TCE crystal, relative peak intensity of S 2s (blue squares, see Fig. 5 b) and electrical anisotropy parameter (see text) - $\Delta R/R$ (closed triangles). Full, dashed and dash-dot lines are to guide the eye, denoting the anisotropy parameter, the Hg and S core level positions, and the intensity ratio of the shifted S 2s HAXPES signal, respectively.

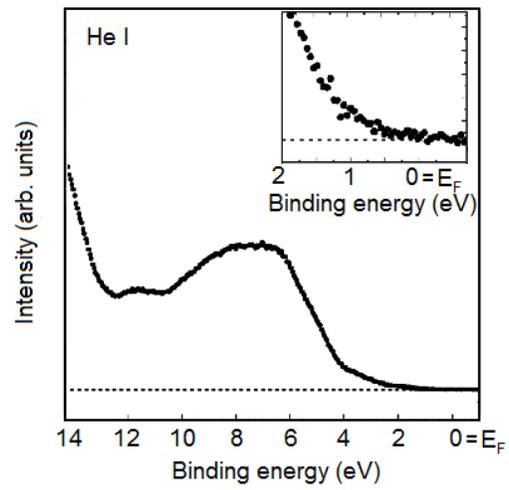

FIG. 7. UPS spectrum recorded at room temperature with a He I source ($h\upsilon = 21.2$ eV). In the inset: blow up of the binding energy region close to the Fermi level.

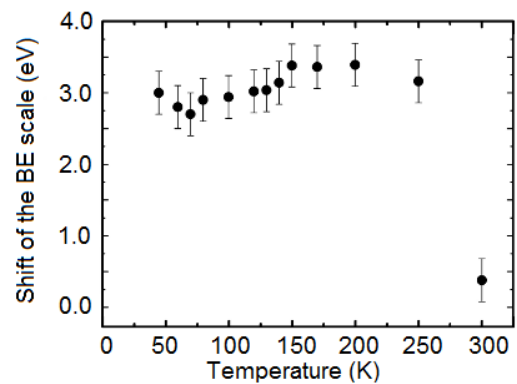

FIG. 8. Temperature dependence of the binding energy (BE) scale shift related to the corresponding He I spectra.

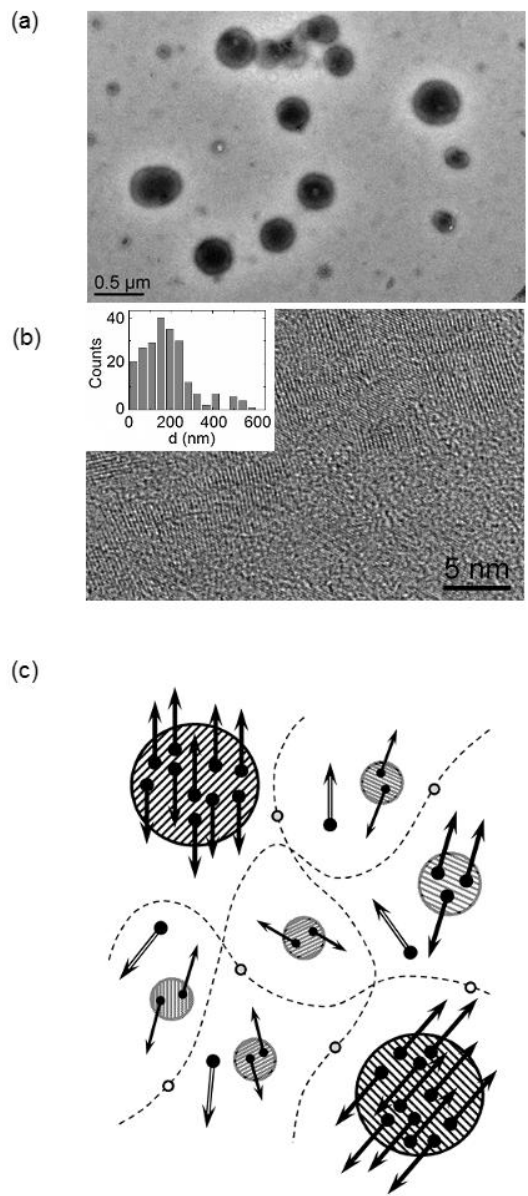

FIG. 9. TEM images of the large (a) and small (b) clusters in the single crystals. The inset shows the size distribution of clusters. (c) Sketch of the objects contributing to the magnetic properties of crystals: antiferromagnetic droplets, individual localized centers and free charge carriers.